\documentclass[a4paper,8pt]{article}
\usepackage[cp1251]{inputenc}
\usepackage[russian,english]{babel}
\usepackage{array,longtable}
\usepackage{amssymb, latexsym, amsthm, amsmath, amsxtra}
\usepackage{graphicx}

\graphicspath{{}}
\DeclareGraphicsExtensions{.pdf,.png,.jpg,}

\makeatletter

\begin{document}
\large

\begin{center}
\title{}{\bf   Young's experiment scheme modification for a possible observation of "soliton" interference model.

 }
\vskip 1cm

\author{}
E.G. Ekomasov \textsuperscript{1} , R.K. Salimov \textsuperscript{1}
{}

\vskip 1cm

{ \textsuperscript{1} Bashkir State University, 450076, Ufa, Russia }


\vskip 0.5cm
e-mail: salimovrk@bashedu.ru

\end{center}

\vskip 1cm

{\bf Abstract}
 \par
We consider the "soliton"  interference model that complements the usual wave and corpuscular models of two-slit interference. The scheme of the experiment to verify such "soliton" interference model has been suggested.

 \par
 \vskip 0.5cm


\par
\vskip 1cm

{\bf 1.Introduction}
\vskip 0.5cm
Nonlinear wave equations, for example Klein-Gordon non-linear differential equations, are the basis of many areas of physics, including hydrodynamics, condensed matter physics, field theory, etc. [1,2] The most studied are (1 + 1)- and (2 + 1)-dimensional models [2-7]. These equations can be easily generalized to higher-dimensional spaces, such as a spherical symmetry:
\begin{align}
  u_{rr}+2 \frac{u_r}{r} -u_{tt}=F(u)
   ,\label{eq:1}
   \end{align}

Stable solutions of these equations can be interpreted as classical models of finite size particles. From this point of view, the space three-dimensional equations are of the main interest. As it is known, there have not been found not splitting (2 + 1) and (3 + 1) oscillating solutions of nonlinear Klein-Gordon equations for the case when the equation has a linear limit while the solution amplitude tends to zero. The case when the equation
\begin{align}
u_{xx}+u_{yy}+u_{zz}-u_{tt}=F(u)
 ,\label{eq:1}
   \end{align}

does not have such limit and the principal member while the solution amplitude tends to zero is non-linear, results in solutions localization in spherically symmetric case. Similar solutions for confining models were considered, for example, in [8] and [9]. These solutions represent not diffusing three-dimensional oscillating solutions. The paper proposed an experiment scheme for the verification of new opportunities for such solutions cooperation.

\vskip 0.5cm

{\bf 2.The Possible "Interference" of the Oscillating Localized Solutions.}

\vskip 0.5cm
Breather-like solutions, for example, equations (3)
 \begin{align}
   u_{rr}+2 \frac{u_r}{r} -u_{tt}=u^\frac{3}{7},\label{eq:1}
   \end{align}
show the constancy of the oscillation fast mode frequency and they are three-dimensional objects[8].

\begin{figure}[!ht]
\center
\includegraphics[width=8cm, height=5cm]{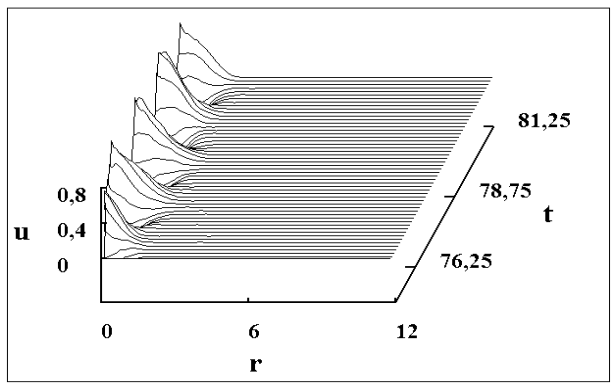}
\caption{Numerical solution of the equation(3) at small times}
\label{schema}
\end{figure}

\begin{figure}[!ht]
\center
\includegraphics[width=8cm, height=5cm]{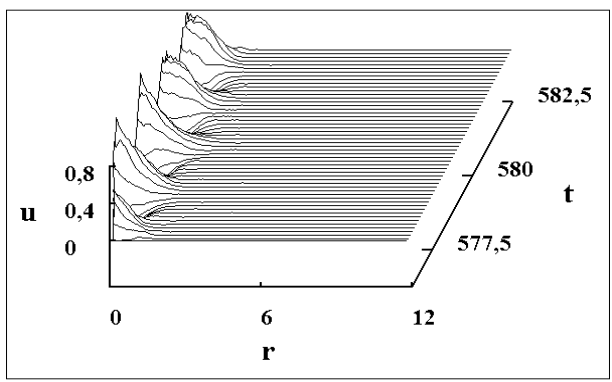}
\caption{Numerical solution of the equation(3) at large times}
\label{schema}
\end{figure}

 These features open new opportunities in the interaction of such objects, for instance, collisions at different angles and wave interference phenomenon. Such Lorentz-invariance breather-like solutions will undergo spatial modulation while moving. Indeed, let us suppose that there is some localized three-dimensional breather-like solution with a preserved rapid fluctuations frequency. Then the solution
\begin{align}
   u(r,t)=\sum  a_i(r,t)cos(\omega_i t),\label{eq:1}
   \end{align}

will be associated with some harmonic function $cos(\omega_j t)$ , where $\omega_j$  - the rapid oscillations mode frequency and coefficient  $a_j$  is substantially bigger than others.
In Lorentz transformation:

\begin{align}
   x'=\frac{x-vt}{\sqrt{1-v^2/c^2}}  ; t'=\frac{t-vx/c^2}{\sqrt{1-v^2/c^2}};y'=y;z'=z      ,\label{eq:1}
   \end{align}

The function $cos(\omega_j t)$  acquire the form
\begin{align}
   cos(\frac{\omega_j(t'-vx')}{\sqrt{1-v^2}});c=1    ,\label{eq:1}
   \end{align}

and the oscillating solution will be spatially modulated in the movement as a de Broglie wave [10]. An example of such transformation is a moving breather in the sine-Gordon equation of the form [2,7]:
\begin{align}
   w(x,t)=4\arctan(\frac{a\sin(b(x-vt)/(\sqrt{1-v^2})}{b\cosh(a(t-vx)/(\sqrt{1-v^2})})   ,\label{eq:1}
   \end{align}

The spatial modulation of such constantly localized three-dimensional solutions should manifest itself in wave interference phenomena – interference. This can be a very interesting situation of "interference" , similar to the hydrodynamic objects interference – the so-called "walkers" in Couder and Fort experiments [11]. Since such oscillating formations tend to localization, the collision of two such objects can also lead to their localization in one state. Indeed, let us consider, for example, the equation:
\begin{align}
   u_{rr}+2 \frac{u_r}{r} -u_{tt}=u^\frac{m}{m},\label{eq:1}
   \end{align}

where $m,n$   - natural numbers of the form $2k+1$,where  $k$ is natural number, besides $m<n$.
For  $v=u/r$, we obtain:
\begin{align}
   v_{rr} -v_{tt}=r^\frac{n-m}{n} v^\frac{m}{m},\label{eq:1}
   \end{align}

This equation shows that the field of the form $ r^\frac{n-m}{n}$ will prevent the diffusion of localized solution. It is natural to assume that the collision of two such solutions leads to their localization in one state. Due to the spatial modulation of the colliding states the movement direction of the resulting state will in a complicated way depend on the direction of the primitive. At the same time taking into account the localization of such oscillating states, the picture of this "interference" will significantly differ from the linear pattern of the unlimited waves interference.
\par
We can view a simple modification of Young experimental scheme, clearly separating these two cases. Fig. 2 shows such a scheme, the localized solutions pass through the diaphragm with two slits, and fall on the screen, here $L$ - is the distance from the diaphragm with two slits to the screen, $d$ - is the distance between the slits, A - is the "soliton" characteristic dimensions. Here, the term "soliton" is referred only to permanent localization of oscillating states, and not their properties during collisions. In scattering on two slits initial localized state or "soliton" 1, passing through two holes splits into two localized "solitons" 2 and 3. Then "solitons" 2 and 3 once again gather in "soliton" 4 similar to 1. Distribution of "soliton" 4 movement directions may create an interference pattern. The wave interference pattern is provided by "solitons" 2 and 3 interaction. Let’s suppose, for simplicity, that the intensity of "solitons" 1 is very little. It is natural to assume that at the proportional increase of $L,d$  sizes, at $d>>A$  , any interference pattern in the "soliton" model should disappear.

\begin{figure}[!ht]
\center
\includegraphics[width=10cm, height=5cm]{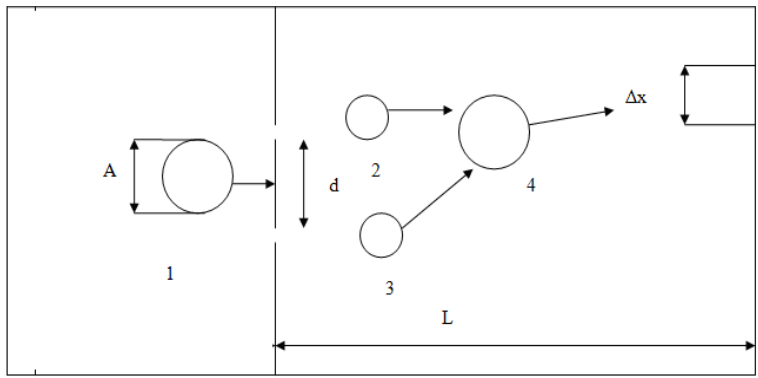}
\caption{ The scheme of a two-slit experiment to verify "soliton" interference pattern}
\label{schema}
\end{figure}

Genuinely, if the "soliton" 1 is localized, it may pass simultaneously through two holes, only if the distance between them is not much bigger than its dimensions. In the case of conventional non-localized waves interference, where:
      \begin{align}
   \frac{\lambda}{\bigtriangleup x}=\frac{d}{L},\label{eq:1}
   \end{align}

proportional increase of $L,d$  sizes  should not lead to a change in the interference pattern, provided maintaining the coherence length and width. Thus, the cases of normal and "soliton" interference can be clearly differentiated. We emphasize once again that according to the authors the "soliton" type of interference is possible only for the solutions which have the property of localization, such as equations solutions of the form (8).
\vskip 0.5cm

{\bf 3.The Possibility of Using Modified Young Experimental Scheme.}
\vskip 0.5cm
For clarity and simplicity, we consider how the proposed experiment scheme will run on the example of particles interference. Young experiment, first conducted in 1801, showed the presence of the light wave properties [12]. The alternative for the Wave theory of light was Newton corpuscular model, where light was presented as a number of particles. Soliton models of particles are, in a sense, intermediate between the concepts of unlimited waves and corpuscles. In this model, it is assumed that the particle is accompanied by some localized wave. An experiment can be carried out to test the ability of "soliton" interference pattern to verify the "soliton" model of any particles.
Let’s consider a particle with mass $m$ , moving at speed $v<<c$ . Assuming that the size of the possible accompanying the particle localized wave is certainly less than $10^{-2}m$, we obtain for $\bigtriangleup x=10^{-4} m$  , $d_0=10^{-2}m$  :
    \begin{align}
d_0= \frac{\lambda L}{\bigtriangleup x}=\frac{h L}{\bigtriangleup x mv}=6.626\cdot 10^{-28}\frac{L}{ mv}  ,\label{eq:1}
   \end{align},
so value $\frac{L}{ mv}$  must be approximately $10^{28}  m^2/J\cdot s$. Assuming that value  L is approximately $1  m$, we obtain an estimate for the value $mv$  of the order  $10^{-28} kg\cdot m/s$. For example, if the particle is an electron, as in Jonson experiment [13], its speed should be approximately 100 m/s. If at greater than $d_0$  distance between the slits $d$ , taking into account the proportional increase of sizes $L,d$  , the interference pattern disappears, then the "soliton" model of particle interference will be demonstrated. In the case of ordinary interference of non-localized waves, the proportional increase in sizes $L,d$  should not lead to a change in the interference pattern. For the real set up of the experiment one can offer such a popular object, as fullerene molecule  $C_{60}$, for which interference experiments has already been conducted (e.g. see [14]).
\vskip 0.5cm
{\bf 4.Conclusions}
\par
\vskip 0.5cm
 In conclusion, we note that although the issue of particles viewed as solitons is very controversial, the "soliton" type of interference may occur in any events. The characteristic feature of "soliton" interference is in the disappearance of the interference pattern on the screen when the distance between the slits is bigger than the value determined by the characteristic dimensions of the "soliton". In conventional interference the proportional increase of $L,d$  parameters does not affect it. Therefore, the proposed experiment scheme to verify the existence of such "interference"  deserves, in the authors opinion, researchers consideration.


\end{document}